\documentclass[runningheads]{llncs}
\usepackage{graphicx}
\usepackage{color}
\usepackage{xcolor}
\usepackage[hidelinks]{hyperref}
\usepackage{mathrsfs}
\usepackage{float}

\newcounter{usesmallsep}
\usepackage{lipsum}

\newcommand{\Fcal}{\mathcal{F}}

\begin{document}
\title{Volume-optimal persistence homological scaffolds of hemodynamic networks covary with MEG theta-alpha aperiodic dynamics
}
\titlerunning{Persistence homological scaffolds of hemodynamic networks}

\author{Nghi Nguyen\inst{1}
\and
Tao Hou\inst{2}\and
Enrico Amico\inst{3}\and
Jingyi Zheng\inst{4}\and 
Huajun Huang\inst{4}\and
Alan D. Kaplan\inst{5}\and 
Giovanni Petri \inst{6}\and
Joaqu\'{i}n Go\~{n}i \inst{7,8}\and
Ralph Kaufmann \inst{9}\and
Yize Zhao \inst{10}\and
Duy Duong-Tran\inst{11,12,13}\and
Li Shen\inst{12,13}
}

\authorrunning{Nguyen \textit{et al.}}

\institute{Multidisciplinary Brain Research Center, Bar-Ilan University, Ramat Gan, Israel 
\and
 Department of Computer Science, University of Oregon, Eugene, Oregon, USA
\and
Institute of Health and Neurodevelopment, College of Health and Life Sciences, Aston University, Birmingham, UK\\ \and
Department of Mathematics and Statistics, Auburn University, Alabama, USA\\ \and
Computational Engineering Division, Lawrence Livermore National Laboratory, Livermore, California, USA\\ \and
NPLab, Network Science Institute, Northeastern University London, London, UK\\ \and
School of Industrial Engineering, Purdue University, West Lafayette, Indiana, USA\\ \and
School of Biomedical Engineering, Purdue University, W. Lafayette, Indiana, USA\\ \and
Department of Mathematics, Purdue University, W. Lafayette, Indiana, USA\\ \and
School of Public Health, Yale University, New Heaven, Connecticut, USA\\ \and
Department of Mathematics, U.S. Naval Academy, Annapolis, Maryland, USA \\ \and 
Department of Biostatistics, Epidemiology and Informatics, University of Pennsylvania, Philadelphia, Pennsylvania, USA\email{\{duyanh.duong-tran,li.shen\}@pennmedicine.upenn.edu}\\\and
Co-supervising Authors
}

\maketitle
\begin{abstract}
Higher-order properties of functional magnetic resonance imaging (fMRI) induced connectivity have been shown to unravel many exclusive topological and dynamical insights beyond pairwise interactions. Nonetheless, whether these fMRI-induced higher-order properties play a role in disentangling other neuroimaging modalities' insights remains largely unexplored and poorly understood. In this work, by analyzing fMRI data from the Human Connectome Project Young Adult dataset using persistent homology, we discovered that the volume-optimal persistence homological scaffolds of fMRI-based functional connectomes exhibited conservative topological reconfigurations from the resting state to attentional task-positive state. Specifically, while reflecting the extent to which each cortical region contributed to functional cycles following different cognitive demands, these reconfigurations were constrained such that the spatial distribution of cavities in the connectome is relatively conserved. Most importantly, such level of contributions covaried with powers of aperiodic activities mostly within the theta-alpha (4-12 Hz) band measured by magnetoencephalography (MEG). This comprehensive result suggests that fMRI-induced hemodynamics and MEG theta-alpha aperiodic activities are governed by the same functional constraints specific to each cortical morpho-structure. Methodologically, our work paves the way toward an innovative computing paradigm in multimodal neuroimaging topological learning. 
The code for our analyses is provided in \url{https://github.com/ngcaonghi/scaffold_noise}.

\end{abstract}
\section{Introduction}
Many real-world systems can be characterized not only by graphs, which are described by nodes and pairwise relations but also by complexes and hypergraphs, elucidating higher-order relations \cite{giusti2015clique,giusti2016two}. Interactions between these higher-order structures have been demonstrated to elucidate complex dynamics, such as phase transitions induced by emergent phenomena in complex networks \cite{battiston2021physics}, that could not be fully explained by pairwise interactions alone. In neuroscience, brain dynamics analysis based on such higher-order descriptions of neuronal populations has yielded statistically and topologically significant results \cite{giusti2015clique,yu2011higher,schneidman2006weak,duong2021morphospace,duong2024homological}. Underlying such successes is a family of analytical tools commonly referred to as Topological Data Analysis (TDA)~\cite{DW22,edelsbrunner2010computational}, which includes persistent homology (PH) \cite{edelsbrunner2000topological,zomorodian2004computing}. Using PH on functional connections deriving from functional magnetic resonance imaging (fMRI), several studies have identified differences between cognitive abilities \cite{ryu2023persistent}, brain states \cite{petri2014homological}, and brain regions \cite{lord2016insights,duong2024homological} in terms of higher-order features (e.g., topological invariants) pertaining to voids (cavities) and cycles. However, the relationships between these topological invariants (e.g., cycles) and other neural correlates, such as anatomical structures and electrophysiological dynamics, remain poorly understood.

Among such neural correlates are cortical aperiodic dynamics, which have recently gained growing interest due to their implicated roles in cognition \cite{ouyang2020decomposing}, consciousness \cite{toker2022consciousness}, aging \cite{voytek2015age}, and diseases \cite{van2023resting,robertson2019eeg,rosenblum2023decreased}. Generative mechanisms of these dynamics, particularly those within the gamma band ($>$ 30 Hz), have been linked to several biological processes \cite{abbas2020geff,abbas2023tangent,chiem2022improving,amico2021toward,garai2023mining,kramer20231,duong2024theorizing}. However, aperiodic activities on the level of cortical regions below the gamma band ($<$ 30 Hz) are significantly less understood.

In this study, leveraging computations in PH, we associate a critical link between fMRI-induced human brain large-scale functional connectivity cycles and aperiodic magnetoencephalography (MEG) activities in theta-alpha range (4-12 Hz). Specifically, we discovered that for each parcellated region in the cortex, changes in its contribution to maintaining homological cycle persistence were positively correlated with changes in its theta-alpha aperiodic power. Comprehensively, our pioneering work paves the way toward a new, innovative computing paradigm in multimodal neuroimaging topological learning.

\section{Methods}

\subsection{Overview of Topological Data Analysis}
Topological Data Analysis (TDA)~\cite{DW22,edelsbrunner2010computational} is an emerging field focusing on understanding the ``shape'' of data (denoted as $X$) through homology theory~\cite{hatcher2002algebraic}. Classical homology groups~\cite{hatcher2002algebraic}  capture voids and cycles of topological spaces in various dimensions $i$ (e.g., connected components ($i=0$), rings ($i=1$), and cavities ($i=2$), and so on). Functional connectomes (FCs, denoted as $G$) in this paper refer to symmetric matrices, whose entries are pair-wise functional couplings quantified by Pearson correlations, realized from $X$. The progression to glean topological information progresses an FC onto a topological space by realizing its simplicial clique complex $\Delta(G)$ which models higher order interactions: $G\rightarrow \Delta(G)$. Defining a threshold $r$, sampled from a set of distinct weight values in FC, determines a corresponding binarized scaffold $G(r)$; the associated homology groups $H_i(\Delta(G))$ becomes a function of $r$. Scanning $r$ from $0$ to $1$ homology is born and annihilated. The sequence of these events is mathematically captured by persistence homology and can be encoded and visualized in terms of bar codes. The topological space, which is simplicial in nature, has associated topological invariants, such as the homology $H_i(\Delta(G))$ and Betti numbers $b_i$. Topological spaces $\Delta(G)$ take the form of a \emph{filtration}
$\Fcal:K_0\subseteq K_1\subseteq\cdots\subseteq K_m$,
which is a growing sequence of discretized spaces called \emph{simplicial complexes}. In $\Fcal$, each simplicial complex $K_i$ is a union of building blocks such as vertices, edges, triangles, and higher dimensional analogs. Given a filtration $\Fcal$, \emph{persistent homology}~\cite{edelsbrunner2000topological} produces a set of intervals (called the \emph{barcodes}) as a scheme for quantifying the ``significance'' (e.g. persistence) of topological features in spaces. In the barcode of $\Fcal$, each interval registers the birth and death of a homology feature in the sequence with the longer intervals typically considered to correspond to more significant features, see~\cite{cohen2007stability,DW22}.

\subsection{Volume-optimal Persistent Cycles}
Homology is the study of classifications of \emph{cycles} in topological spaces and features in homology are also represented by the cycles. However,  the barcode of a filtration $\Fcal$ does not provide concrete representative cycles for the homology features born and died in the sequence. These representatives for persistent homology, termed as \emph{persistent cycles}, turned out to reveal crucial insight into the topological spaces over which filtrations are taken.

\begin{definition}
For an interval $[b,d)$ in the barcode of $\Fcal$, a \emph{persistent cycle} for $[b,d)$ is a cycle born in $K_b$ (existing in $K_b$ but not in $K_{b-1}$) and becoming trivial in $K_d$ (is not a boundary in $K_{d-1}$ but becomes  a boundary in $K_d$).
\end{definition}

Notice that boundaries are special cycles  representing the trivial homology feature (zero). As indicated in several works (e.g.,~\cite{dey2020computing,obayashi2018volume}), an interval in the barcode could have more than one  persistent cycles, among which a canonical choice is to compute an \emph{optimized} persistent cycle with weights typically reflecting the intrinsic areas of the cycles. Such weight-optimized cycles can provide the tightest representations of the holes characterized by persistent homology. However, Dey et al.~\cite{dey2020computing} showed that computing optimal persistent cycles is NP-hard in general. In this paper, we utilize a Python library called \texttt{HomCloud} developed by Obayashi~\cite{obayashi2018volume} for computing persistent cycles.
\texttt{HomCloud} uses a linear-program-based algorithm for computing optimized persistent cycles  and indeed targets  \emph{volume-optimal} persistent cycles, where weights of the higher dimensional bounding chains are optimized.  In this paper, we found that the persistent cycles produced by \texttt{HomCloud} typically have good quality (better than the non-optimized ones produced by the original persistence algorithm \cite{edelsbrunner2000topological}). More importantly, mathematical constructs based on such cycles can afford additional interpretations about the volume of the cavities that they wrap around.

\subsection{HCP data processing}

    \subsubsection{Persistence homological scaffolds of resting-state and task-based hemodynamic networks.}
    
        We used resting-state and task-based 3T fMRI data from a subset of the Human Connectome Project Young Adult dataset \cite{van2013wu}, preprocessed by the Neuromatch Academy \cite{t2022neuromatch}. Our study focused on left-hand, right-hand, left-foot, and right-foot motor tasks, and 0-back faces, 0-back tools, 2-back faces, and 2-back tools working memory tasks. To ensure valid extrapolation to MEG neural correlates, we excluded 20 subjects that underwent the same task conditions during MEG recording trials. The resulting sample size was N = 319, each of which represents the BOLD signals of one subject under resting and task conditions and parcellated according to the Glasser atlas \cite{glasser2016multi}. 

    \begin{figure}[ht!]
      \centering
      \includegraphics[width=\columnwidth]{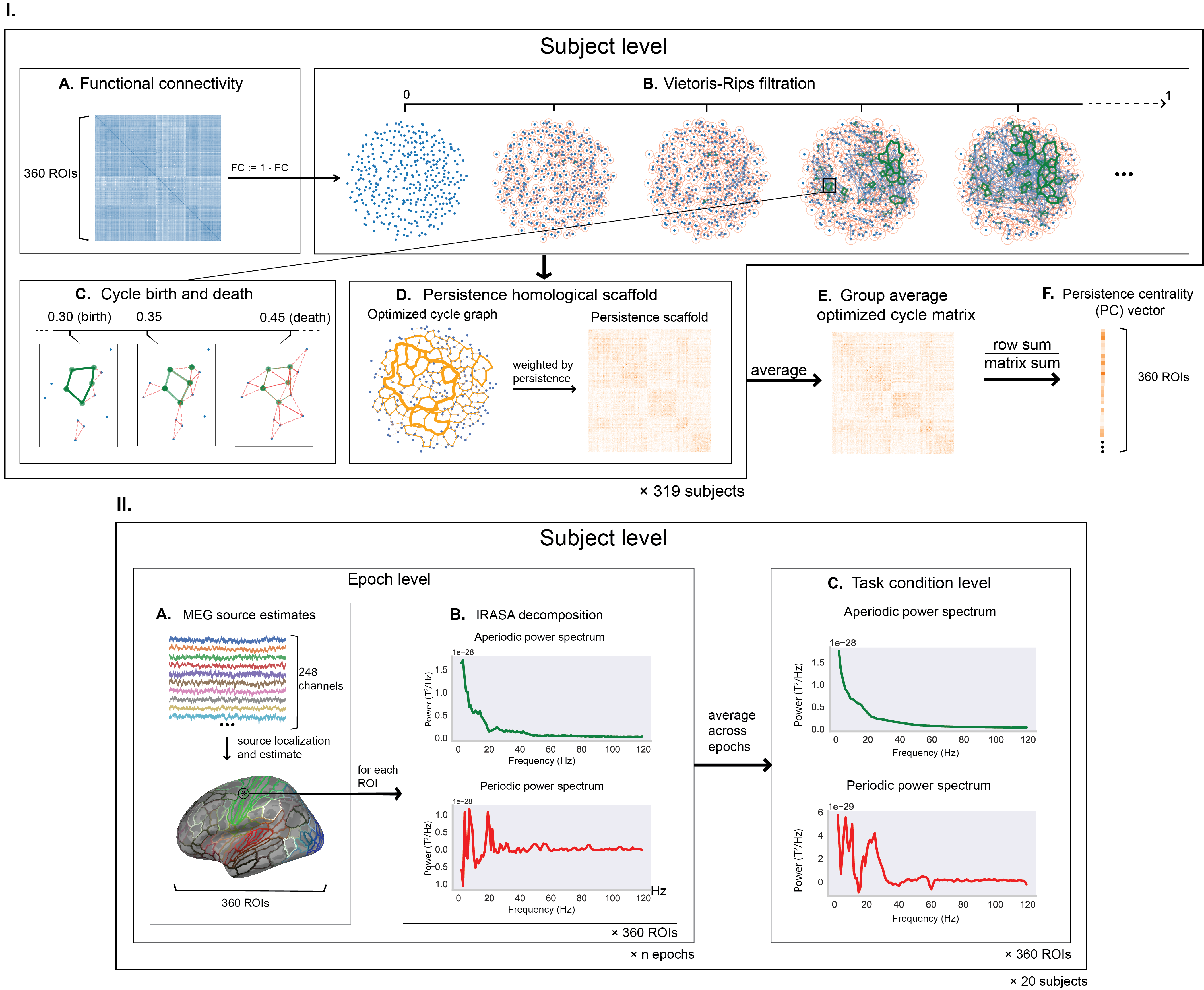}
      \caption{
      Data processing pipelines. I. From fMRI functional connectivity to persistence centrality. II. From sensor-level MEG signals to Glasser-parcellated power spectrum decomposition. 
    }
      \label{fig:resultfig1}
    \end{figure}

        Functional connectivity matrices were computed for each subject by computing Pearson correlation coefficients between BOLD time series of region pairs. To perform persistent cycle discovery with \texttt{HomCloud} on every connectivity matrix $G$, we converted each pairwise correlation term $r$ into a pseudo-distance index $\omega := 1 - r$ while excluding negative correlations (figure 1.I.A). For each pseudo-distance matrix with weights $\{\omega\}$, we generated a Vietoris-Rips filtration $\mathcal{F}$ of the homological group $H_1$. Every complex in $\mathcal{F}$ entailed a birth-death interval $[b, d)$ where $b, d \in \{\omega\}$, a volume-optimal persistent cycle $g_t$, and its persistence $\pi_{g_t} = d - b$ (figures 1.I.B and 1.I.C). From this point, we followed the approach outlined by Petri \textit{et al.} ~\cite{petri2014homological} to calculate the \emph{persistence homological scaffold network} $\mathcal{H}^p_{G}$ composed of the volume-optimal cycles (figure 1.I.D). Specifically, for every edge $e_{i, j}$ between nodes $i$ and $j$ belonging to multiple cycles $g_0, g_1, \cdots, g_s$, its weight is
        $
            w_{i, j}^{\pi} = \sum_{g_{t} \mid e_{i, j} \in g_{t}} \pi_{g_{t}}.
        $
        The group-averaged scaffold matrix was generated for each experimental condition (figure 1.I.E), from which we calculated the \emph{persistence centrality} (PC) indices defined by
        $
            PC(i) =  \sum_{j}w^{\pi}_{i, j}\left(\sum_{j, k}w^{\pi}_{j, k}\right)^{-1}
        $
        \noindent for each node $i$ (figure 1.I.F). For comparative purposes, we also calculated the degree centrality (DC) indices from the original functional connectivity matrix $G$ with edge weights $\{r\}$, whereby
        $
            DC(i) =  (\sum_{j}r_{i, j})(\sum_{j, k}r_{j, k})^{-1}.
        $

    \subsubsection{MEG data preprocessing.}
    We used MEG data from the same HCP dataset under the same task conditions but with the subjects previously excluded from the fMRI data analysis (n = 20) and preprocessed as previously detailed by Larson-Prior et al., 2013 \cite{larson2013adding}. Each subject's data comprised 12-second segments of randomly epoched resting-state data and task-evoked data starting from the first stimulus onset of each task block. We performed source reconstruction using linearly constrained minimum variance (LCMV) beamformers \cite{van1997localization} on boundary element method (BEM) surfaces derived from subject anatomy data. The source estimates were then grouped by labels defined by the Glasser atlas and averaged across sources (figure 1.II.A), resulting in a single time series per epoch for every Glasser region.

\subsubsection{Power spectrum decomposition.}
    On each Glasser-parcellated MEG epoch, we computed the power spectral density using Welch's method (with 2-second Hann windows and 1-second overlaps) and performed spectrum decomposition using the irregular-resampling auto-spectral analysis (IRASA) method developed by Wen \& Liu, 2015 \cite{wen2016separating}. The resulting aperiodic and periodic spectra spanned a frequency band of 1 to 120 Hz (Figure 1.II.B). Epoch-level spectra were averaged to produce those for the resting state and every task condition (Figure 1.II.C). To avoid high-frequency artifacts that might contaminate downstream analyses, we only considered power in the range of 1 to 90 Hz.

\section{Result}

\subsection{Directions of change in persistence centralities are consistent across attentional task conditions}
\begin{figure}[ht!]
  \centering
  \includegraphics[width=\columnwidth]{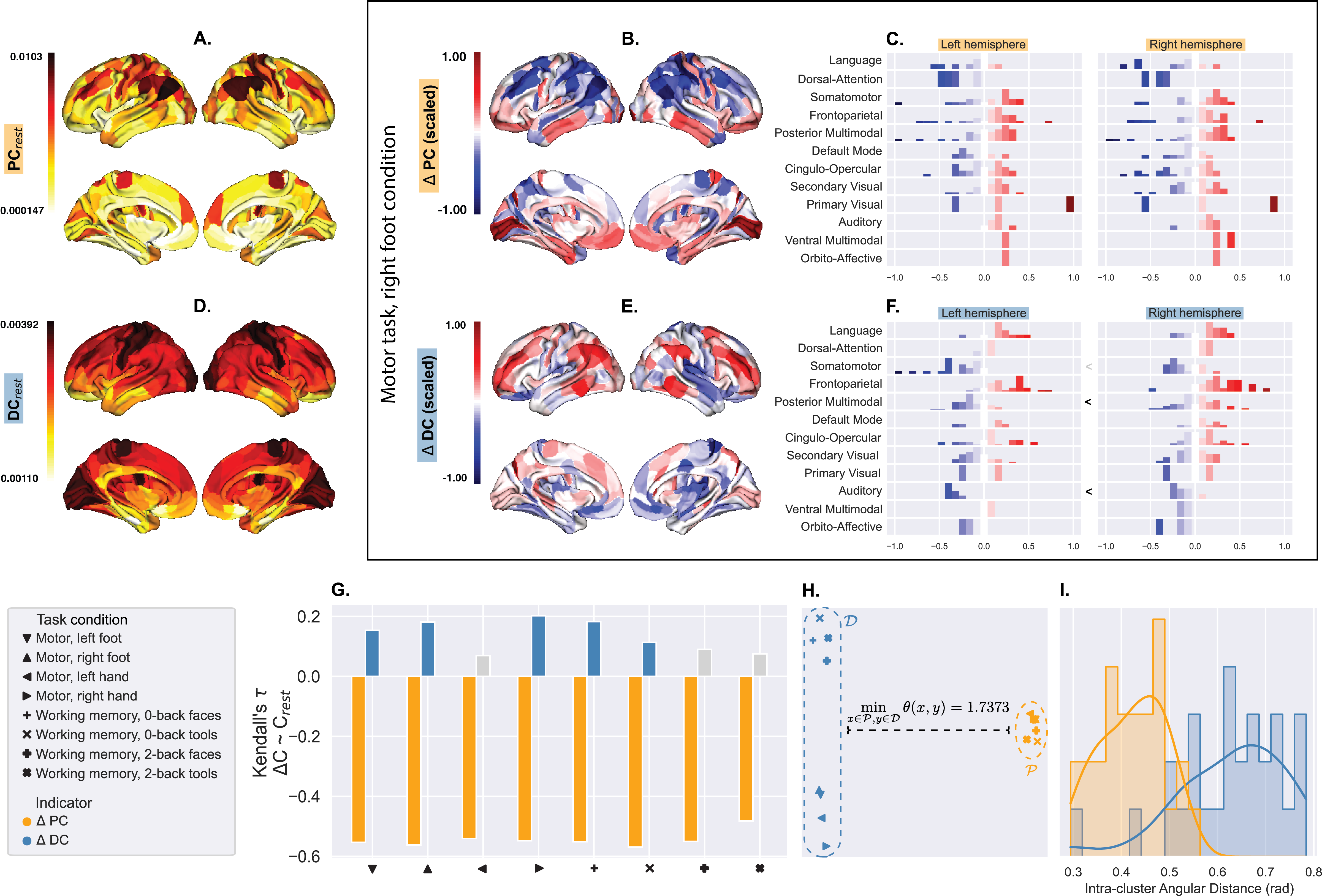}
  \caption{
  Resting state-subtracted PC ($\Delta PC$) captures task-evoked hemodynamic patterns nonspecific to task conditions, whereas resting state-subtracted DC ($\Delta DC$) is more condition-specific.
}
  \label{fig:resultfig3}
\end{figure}
In Figure 2.A and 2.D, the spatial distributions of resting-state PC and DC are portrayed on a Glasser-parcellated surface. Figure 2.B maps $\Delta PC$, representing the PCs during the motor task and right foot condition subtracted by the resting-state PCs, on the same brain surface. The mapped surface shows a coarse-grained global gradient of $\Delta PC$ starting from the deepest $\Delta PC$ sinks (in blue) in the dorsal areas to the sources (in red) in ventral regions. Notably, this gradient does not exhibit a hemisphere asymmetry (figure 2.C), as evidenced by similar distributions of sinks and sources across hemispheres for each functional network ($p > 0.34 > 0.05$, Mann-Whitney $U$-test). Figures 2.E and 2.F demonstrate the $\Delta DC$ counterparts of figures 2.B and 2.C, respectively, revealing a different distribution of sinks and sources that shows hemispheric asymmetry. See Supplementary information, figures S1 and S2 for a comprehensive comparison of $PC$, $DC$, $\Delta PC$, and $\Delta DC$ distributions across all task conditions.

Figure 2.G demonstrates the relationship between resting-state $PC$ (or $DC$) and $\Delta PC$ (or $\Delta DC$) evoked by each task condition. All $\Delta PC$ vectors have strongly negative correlations with the resting-state $PC$ ($p << 0.05$, Kendall's $\tau$ for rank association), while their $\Delta DC$ counterparts show weak or non-significant correlations. $\Delta PC$ and $\Delta DC$ clusters are at least not positively correlated as demonstrated by the minimum angular difference of $1.7373$ radians, \textit{i.e.}, nearly orthogonal (visualized by multidimensional scaling, MDS, in figure 2.H). Figure 2.I sketches the distributions of angular differences within each cluster, showing that spatial patterns of $PC$ sinks and sources are more consistent across tasks and conditions than those of $DC$.

From the observations above, we concluded that $\Delta PC$ was more correlated with changes in cognitive demand than with the nature of the task, while $\Delta DC$ was task-specific. Spatially, $\Delta PC$ exhibited patterns such as ``sinks" concentrated in the Dorsal-Attention network and ``sources" in the Auditory network and part of the Cingulo-Opercular network previously identified as the salience network \cite{ji2019mapping}. Such ``sink" and ``source" regions indicated the reduction and increase, respectively, of the volumes of functional connectome cavities incident to those regions. Since the spatial distributions of $\Delta PC$ ``sinks" and ``sources" were consistent across task conditions (Supplementary information, figures S1), we further concluded that under different cognitive demands, cavities in the functional connectomes were conserved in space and relatively invariant to changes in cortical computations.

\subsection{Persistence centralities and theta-alpha aperiodic band powers change in the same direction}
\begin{figure}[h!]
    \centering
    \includegraphics[width=\columnwidth]{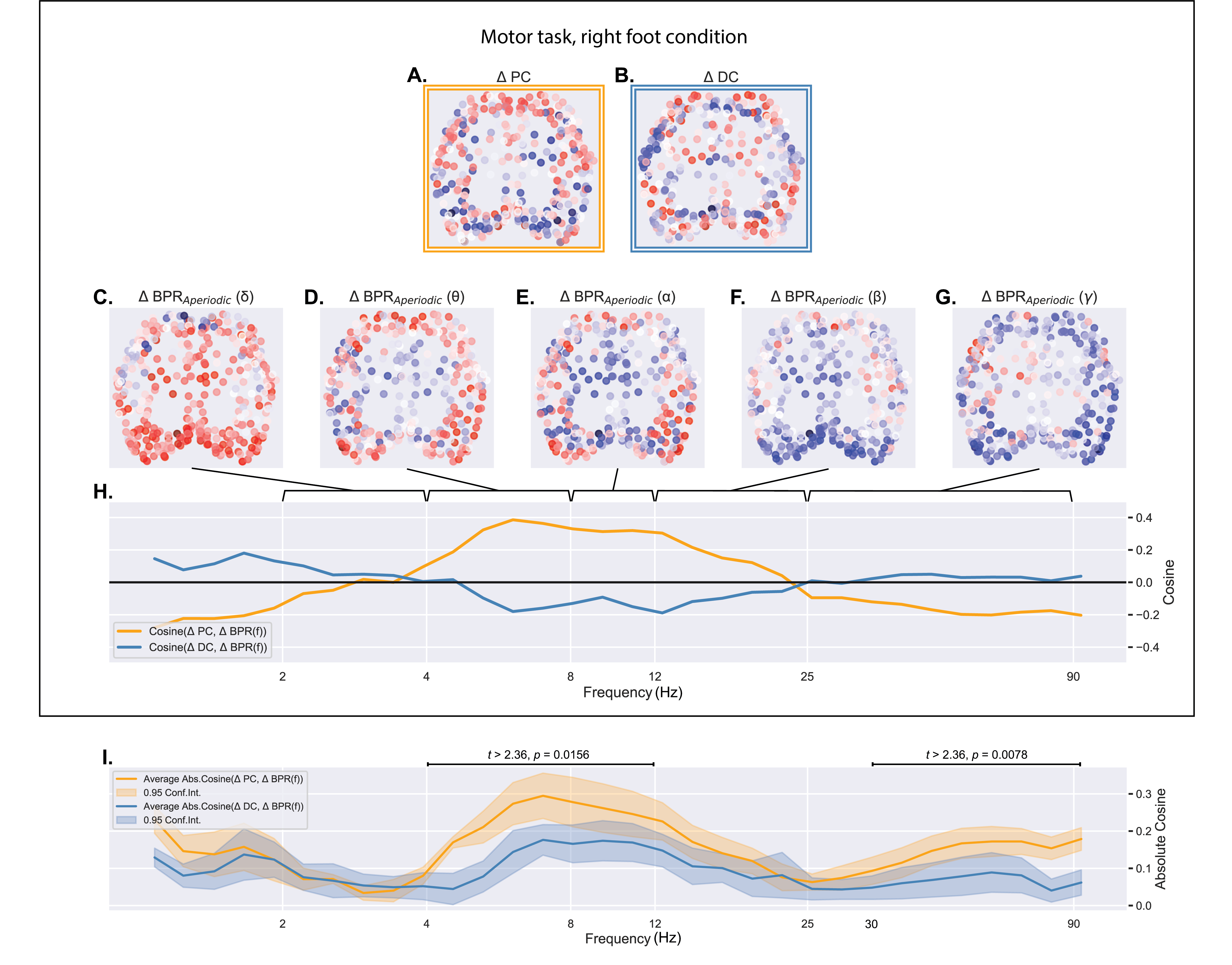}
    \caption{Regional alterations in bandpower ratio ($\Delta$ BPR) of cortical aperiodic activity within the theta-alpha range (4 - 12 Hz) exhibit a stronger alignment with $\Delta PC$ compared to $\Delta DC$. }
    \label{fig:resultfig4}
\end{figure}

We observed that $\Delta PC$ coincided with aperiodic dynamic patterns outlined in previous fMRI-EEG studies \cite{jacob2021aperiodic}. Further corroborating this correlation, cosine similarities between $\Delta PC$ and changes in band power ratios ($\Delta BPRs$) within the theta-alpha range were significantly higher than those between $\Delta BPRs$ and $\Delta DC$.

Figures 3.A and 4.B present Isomap projections of $\Delta PC$ and $\Delta DC$, respectively, evoked by right-foot cues during the motor task. These allow for visual comparison with the $\Delta$ BPR of canonical bands (delta: 2-4 Hz, theta: 4-8 Hz, alpha: 8-12 Hz, beta: 12-25 Hz, gamma: 25-90 Hz), also projected using Isomap and depicted in figures 3.C to 4.G. $\Delta$ BPR was also computed for 32 frequency bands whose endpoints were logarithmically spaced within the range of 2 to 90 Hz. Cosine similarities between these $\Delta$ BPRs and $\Delta PC$ (orange) or $\Delta DC$ (blue) under the motor, right-foot condition are shown in Figure 3.H. Absolute cosine similarities averaged across conditions are shown in Figure 3.I, suggesting that $\Delta PC$ better predicts $\Delta$ BPRs in the theta-alpha and gamma ranges (p = 0.0156 and p = 0.0078, respectively, cluster-based permutation paired t-test). Similar analyses for the periodic component resulted in statistically insignificant correlations between periodic activities and $\Delta PC$ or $\Delta DC$.

\section{Discussion}
By exploring the relationship between fMRI and MEG data from a novel bi-modal perspective, this study reveals two previously undocumented phenomena. Firstly, the persistence centrality (PC) of each cortical structure, representing its participation in persistent cycles relative to the total participation, demonstrates predictable changes between brain states induced by different attentional demands. Specifically, regardless of the task nature, functional cavities incident to structures with high PCs tend to contract when attention is required, and vice versa. Secondly, the directions of change in PC predict those in the power of theta-alpha aperiodic activities on the brain-wide scale. These observations underscore a relationship between BOLD signals and electrophysiological activities that cannot be described or explained only in terms of pairwise interactions captured by node degrees. In other words, it must rely on the notion of volume-optimal functional cycles. 

Since changes in these functional cycles and their associated changes in the aperiodic spectra are robust across normal brain states, it might be possible that they emerge from a common constraint regime governing both hemodynamics and aperiodic electrodynamics. This relatively new notion of a constraint regime described by cycles suggests new directions toward understanding the generative mechanisms of BOLD signals and sub-gamma aperiodic spectra. Future investigations---which might involve connections to subcortical structures such as the thalamus---can help detail or expand the current models of BOLD functional connectivity on the systemic macrovascular level. More broadly, by leveraging topological assumptions that persist throughout different normal brain states, homological cycles in the brain can provide better guidance for treatments of outlying brain states indicative of psychological disorders, neuropathologies, and neurodegeneration \cite{xu2022consistency,xu2024topology}. 

\begin{credits}
\subsubsection{\ackname} 
LS acknowledges financial support from the National Institutes of Health grants RF1 AG068191, U01 AG066833, and U01 AG068057. DDT acknowledges financial support from the Office of Naval Research N0001423WX00749. NN acknowledges financial support from the Erasmus Mundus Joint Master's Programme in Brain and Data Science, European Commission. TH acknowledges financial support from the National Science Foundation grant CCF 2348238. Data used in this project were obtained from the Human Connectome Project, WU-Minn HCP data – 1200 subjects (db.humanconnectome.org).

\subsubsection{\discintname}
The authors have no competing interests to declare that are
relevant to the content of this article.
\end{credits}

\bibliographystyle{splncs04}
\bibliography{Paper}
\end{document}